\documentclass[pdflatex,sn-mathphys-num]{sn-jnl}% Math and Physical Sciences Numbered Reference Style 
\usepackage{graphicx}%
\usepackage{multirow}%
\usepackage{amsmath,amssymb,amsfonts}%
\usepackage{amsthm}%
\usepackage{mathalpha}
\DeclareMathAlphabet{\mathscr}{U}{rsfs}{m}{n}
\usepackage[version=4]{mhchem}
\usepackage[title]{appendix}%
\usepackage{xcolor}%
\usepackage{textcomp}%
\usepackage{manyfoot}%
\usepackage{booktabs}%
\usepackage{algorithm}%
\usepackage{algorithmicx}%
\usepackage{algpseudocode}%
\usepackage{listings}%
\usepackage{subcaption}
\usepackage{tabularx} % For resizing tables
\usepackage{lmodern}

\begin{document}

\title[Article Title]{Scaling active spaces in simulations of surface reactions through sample-based quantum diagonalization}

%%=============================================================%%
%% GivenName	-> \fnm{Joergen W.}
%% Particle	-> \spfx{van der} -> surname prefix
%% FamilyName	-> \sur{Ploeg}
%% Suffix	-> \sfx{IV}
%% \author*[1,2]{\fnm{Joergen W.} \spfx{van der} \sur{Ploeg} 
%%  \sfx{IV}}\email{iauthor@gmail.com}
%%=============================================================%%

\author[1,6]{\fnm{Marco Antonio} \sur{Barroca}}%\email{mabarroca@ibm.com}

\author[2]{\fnm{Tanvi} \sur{Gujarati}}%\email{tgujarati@ibm.com}

\author[3]{\fnm{Vidushi} \sur{Sharma}}%\email{vidushis@ibm.com}

\author[1]{\fnm{Rodrigo} \sur{Neumann Barros Ferreira}}%\email{rneumann@br.ibm.com}

\author[3]{\fnm{Young-Hye} \sur{Na}}%\email{yna@us.ibm.com}

\author[3]{\fnm{Maxwell} \sur{Giammona}}%}\email{Maxwell.Giammona@ibm.com}

\author[4]{\fnm{Antonio} \sur{Mezzacapo}}%\email{mezzacapo@ibm.com}

\author[5]{\fnm{Benjamin} \sur{Wunsch}}%\email{bhwunsch@us.ibm.com}

\author*[1]{\fnm{Mathias} \sur{Steiner}}\email{mathiast@br.ibm.com}

\affil[1]{\orgname{IBM Research}, \orgaddress{\city{Rio de Janeiro}, \postcode{20031-170}, \state{RJ}, \country{Brazil}}}

% \affil*[1]{\orgdiv{Department}, \orgname{IBM Research}, \orgaddress{\street{Street}, \city{City}, \postcode{100190}, \state{State}, \country{Country}}}

\affil[2]{\orgname{IBM Quantum}, \orgname{IBM Research - Almaden}, \orgaddress{\city{San Jose}, \postcode{95120}, \state{CA}, \country{USA}}}

\affil[3]{\orgname{IBM Research}, \orgname{IBM Research - Almaden}, \orgaddress{\city{San Jose}, \postcode{95120}, \state{CA}, \country{USA}}}

\affil[4]{\orgname{IBM Quantum}, \orgname{IBM T.J. Watson Research Center, Yorktown Heights}, \orgaddress{\city{New York}, \postcode{10598}, \state{NY}, \country{USA}}}

\affil[5]{\orgname{IBM Research}, \orgname{IBM T.J. Watson Research Center, Yorktown Heights}, \orgaddress{\city{New York}, \postcode{10598}, \state{NY}, \country{USA}}}

\affil[6]{ \orgname{Centro Brasileiro de Pesquisas Físicas}, \orgaddress{\city{Rio de Janeiro}, \postcode{22290-180}, \state{RJ}, \country{Brazil}}}

%%==================================%%
%% Sample for unstructured abstract %%
%%==================================%%

\abstract{Quantum-chemical simulations are essential for predicting energies of chemical reactions. Accurately solving the many-body Schr\"{o}dinger equation for reagent and product states of most relevant chemical process is, however, unfeasible. Quantum computing offers a pathway for predicting energies of correlated electronic systems with localized interactions. Here, we apply a quantum embedding approach for investigating oxygen reduction reactions at the electrode surface in Lithium batteries, a representative example of energetic analysis in localized chemical reactions. We employ an Active Space Selection method based on Density Difference Analysis for identifying the orbitals involved in the reaction. Leveraging the Local Unitary Cluster Jastrow ansatz for state preparation, the active-space orbitals are then processed on a quantum computer. As quantum algorithms, we use Sample-based Quantum Diagonalization, SQD, and its extended version, Ext-SQD, which integrates electronic excitations into the quantum-selected electronic configuration subspace. The largest configurations are represented by quantum circuits mapped onto 80 qubits of an IBM Heron R2 quantum processing unit. For up to 12 orbitals, we are able to benchmark the quantum-computed reaction energies against results obtained with Complete Active Space Configuration Interaction. For benchmarking results in active spaces as large as 32 orbitals, we resort to Heat-Bath Configuration Interaction and Coupled Cluster Singles and Doubles calculations, respectively. At 27 orbitals, the Ext-SQD results exhibit prediction accuracy improvements with regard to the standard, quantum-chemical reference methods that remain computationally feasible at that scale. The results indicate the potential of sample-based quantum diagonalization for performing high-accuracy reaction modeling in chemistry and materials science.}


%%\pacs[JEL Classification]{D8, H51}

%%\pacs[MSC Classification]{35A01, 65L10, 65L12, 65L20, 65L70}

\maketitle

\section*{Introduction}\label{sec:Int}

Computational modeling of correlated electronic systems is essential for accurately predicting chemical reaction energies. Exact solutions of the electronic Schrödinger equation are unfeasible, forcing the adoption of approximate methods. Although Density Functional Theory (DFT) is widely used, its exchange-correlation approximations fail to capture strong electron-electron correlations in complex systems \cite{Cohen2008, Jones1989}.

%To overcome these challenges, hybrid approaches that combine classical and quantum computing resources are being actively explored. This essentially means quantum computers must be used alongside current high-performance computing hardware and classical techniques in what is called Quantum-centric Supercomputing.\cite{Alexeev2024-ms} Quantum computing offers a fundamentally different paradigm for solving electronic structure problems, leveraging quantum algorithms that, in principle, scale more favorably than classical methods. Techniques such as Quantum Phase Estimation (QPE) have been proposed to simulate electronic wavefunctions more efficiently.\cite{Kang2022, cortes2024assessingquerycomplexitylimits, Goings2022} However, current quantum hardware limitations require hybrid strategies that integrate quantum methods with classical simulations to optimize performance and scalability as it is not yet viable to perform full electronic structure calculations on quantum hardware alone.\cite{Clary2023-gz}

Hybrid quantum-classical approaches, also termed Quantum-centric Supercomputing \cite{Alexeev2024-ms}, offer an alternative pathway by leveraging quantum algorithms that, in principle, simulate electronic wavefunctions more efficiently than classical methods~\cite{Kang2022, cortes2024assessingquerycomplexitylimits, Goings2022}. Due to today's quantum hardware limitations, short-term approaches to practical quantum computing in chemistry integrate quantum algorithms within classical simulation workflows. An important application is the simulation of select active orbital spaces, a technique which reduces the size of the system under study while preserving its characteristic electronic correlations. Demonstrated approaches include Density Difference Analysis (DDA) with coupled-cluster natural orbitals \cite{Gujarati_2023}, Atomic Valence Active Space (AVAS) with atomic valence orbitals \cite{Sayfutyarova2017-rl}, and Density Matrix Embedding Theory (DMET) with partitions of correlated fragments \cite{Sun2014-bq}.

The Variational Quantum Eigensolver (VQE)~\cite{Peruzzo2014, Reiher2017, Wecker2014} method has been explored for predicting molecular ground states as well as reaction energies. However, challenges in ansatz design, the occurrence of barren plateaus and a high sampling penalty have posed practical limitations to the size of the chemical system under study \cite{Fedorov2022-pp, Clary2023-gz, Holmes2022, Singkanipa2025-lm}. Recently, Sample-Based Quantum Diagonalization (SQD) was successfully applied to project an electronic Hamiltonian onto a reduced subspace constructed from a sampled set of electronic configurations, before processing the quantum-selected subspace with classical diagonalization techniques\cite{robledomoreno2024chemistryexactsolutionsquantumcentric}. An enhanced version of this method, Extended SQD (Ext-SQD), introduces excitation operators to refine the configuration space and improve the accuracy of energy predictions \cite{barison2024quantumcentriccomputationmolecularexcited}. In the SQD framework, a process termed configuration recovery plays a crucial role. By iteratively correcting the sampled electronic configurations with regard to electron number and symmetry constraints, the recovery step validates the configurational outcomes.

For constructing a reduced subspace to be used with SQD and Ext-SQD, the electronic configurations must be efficiently sampled with a quantum processing unit. The Local Unitary Cluster Jastrow (LUCJ) ansatz enables a minimum gate depth while preserving electronic correlations. Initializing circuits with coupled-cluster parameters taken from "classical", i.e., quantum-chemical simulation methods performed with standard computers,  further enhances the quality of selected configurations \cite{Motta_2023}. The approach enables accurate predictions of ground-state and excited-state energies, allowing for the quantum-assisted predictions to be scaled up and to be benchmarked against established reference methods of quantum chemistry \cite{liepuoniute2024quantumcentricstudymethylenesinglet, robledomoreno2024chemistryexactsolutionsquantumcentric, barison2024quantumcentriccomputationmolecularexcited, Motta_2023}.

In this work, we apply an advanced, hybrid quantum-classical method based on SQD and Ext-SQD for predicting the ground state energy difference of reagent and product states in molecular surface reactions. We increase stepwise the orbital count in the active space and sample the most relevant electronic configurations by means of a customized LUCJ circuit on a pre-fault tolerant quantum device. This way, we are able to benchmark the prediction accuracy obtained with the quantum-sampled configurations against the results obtained with high-accuracy, quantum-chemical simulations performed on standard computers.

% The reaction energy, defined as the energy difference $\Delta E = E_{\text{prod}} - E_{\text{reac}}$ between the reactant and product states, provides insights into the thermodynamics of lithium oxide formation-a crucial aspect of battery stability. To study this system, we employed a computational workflow consisting of geometry optimization, active space selection, and quantum experiments. This workflow, summarized in \autoref{fig:AS_worflow}, enables a systematic investigation of lithium-oxygen interactions at the atomic level.

% Our results highlight the potential of integrating quantum computing with classical embedding techniques for advancing materials discovery and reaction modeling. By systematically selecting active spaces and leveraging sample-based diagonalization methods, we provide a scalable and effective approach for simulating complex chemical systems. In particular, we demonstrate that incorporating excitation operators in Ext-SQD significantly enhances the quality of sampled configuration spaces compared to standard SQD, leading to improved accuracy in representing ground state wavefunctions without entirely sacrificing computational efficiency. Finally, we discuss the broader implications of our work and propose directions for future research to further refine and scale quantum-assisted simulations in chemical and materials science.

Specifically, we show that incorporating excitation operators into SQD increases the quality of sampled configurations, thus improving prediction accuracy without sacrificing computational efficiency. Using a scalable SQD workflow, we execute large circuits of up to 80 qubits on \texttt{ibm\_aachen}, a quantum computer equipped with a Heron R2 processing unit containing 156 qubits. Starting from DFT-based atomic geometries and energies, we find that Ext-SQD matches and, at larger active spaces, eventually outperforms classical simulation methods. To our knowledge, this is the first application of SQD and Ext-SQD in chemical reaction predictions. Finally, we discuss the implications of our study and outline next steps to further refine and scale quantum-assisted simulations in chemistry and materials science.

% \begin{figure}[!htb]
% \centering
% \includegraphics[width=0.85\textwidth]{imgs/workflow_2.png}
% \caption{\textbf{Hybrid quantum-classical framework for studying lithium–oxygen reactions.}  Geometry optimization and DFT calculations lead into the Active space selection via DDA and Natural Orbitals which identifies the key orbitals, and a LUCJ ansatz samples configurations from quantum hardware. SQD/Ext-SQD then compute reaction energies ($\Delta E = E_{\text{prod}} - E_{\text{reac}}$), providing insights into lithium oxide formation.} \label{fig:AS_worflow}
% \end{figure}

\section*{Results}\label{sec:Res}

\subsection*{Surface Reaction}

The reaction of lithium and oxygen in Li-metal anode batteries plays a crucial role in stabilizing the solid electrolyte interphase (SEI) and enhancing battery performance. While offering high energy density, Li-batteries face significant safety challenges due to dendrite formation, which can lead to short circuits and failure. The inclusion of dissolved oxygen in the electrolyte has been shown to mitigate dendrite growth by promoting the formation of lithium oxides, thereby improving stability and cycle life. \cite{Giammona2023}

The primary reaction of interest in these systems is \[\ce{4 Li + O2 -> 2 Li2O},\] resulting in the formation of lithium oxide (\ce{Li2O}), which is deposited at the SEI. However, depending on local conditions such as oxygen concentration and electrolyte composition, other lithium oxides, including lithium peroxide (\ce{Li2O2}) and lithium superoxide (\ce{LiO2}), may also form. The competition between these different reaction pathways is a critical factor influencing battery performance and longevity. \cite{Giordani2013-ke, Zheng2022}

Previous studies have proposed reaction mechanisms for various surface-mediated electrochemical processes~\cite{Ince2023, Liu2023,Di_Paola_2024}. Following a similar approach, we investigated the formation of \ce{Li+} and \ce{O2-} species on the surface through quantum chemistry calculations. The reaction of lithium with oxygen involves the dissociation of molecular \ce{O2} and the subsequent adsorption of \ce{O2-} anions, which interact with \ce{Li+} ions to form stable surface-bound lithium oxides. This process can be described by \[\ce{Li + O2 -> Li+ + O2-(ads)}.\]

The adsorption of \ce{O2-} plays a crucial role in the initial stages of lithium oxide formation. The stabilization of the anions on the surface dictates the pathway for further oxide growth, ultimately influencing the electrochemical performance of the battery. \cite{Giordani2013-ke,Shen2018-ge, Wang2020-zl} 

\begin{figure}[!htb]
\centering
\includegraphics[width=0.99\textwidth]{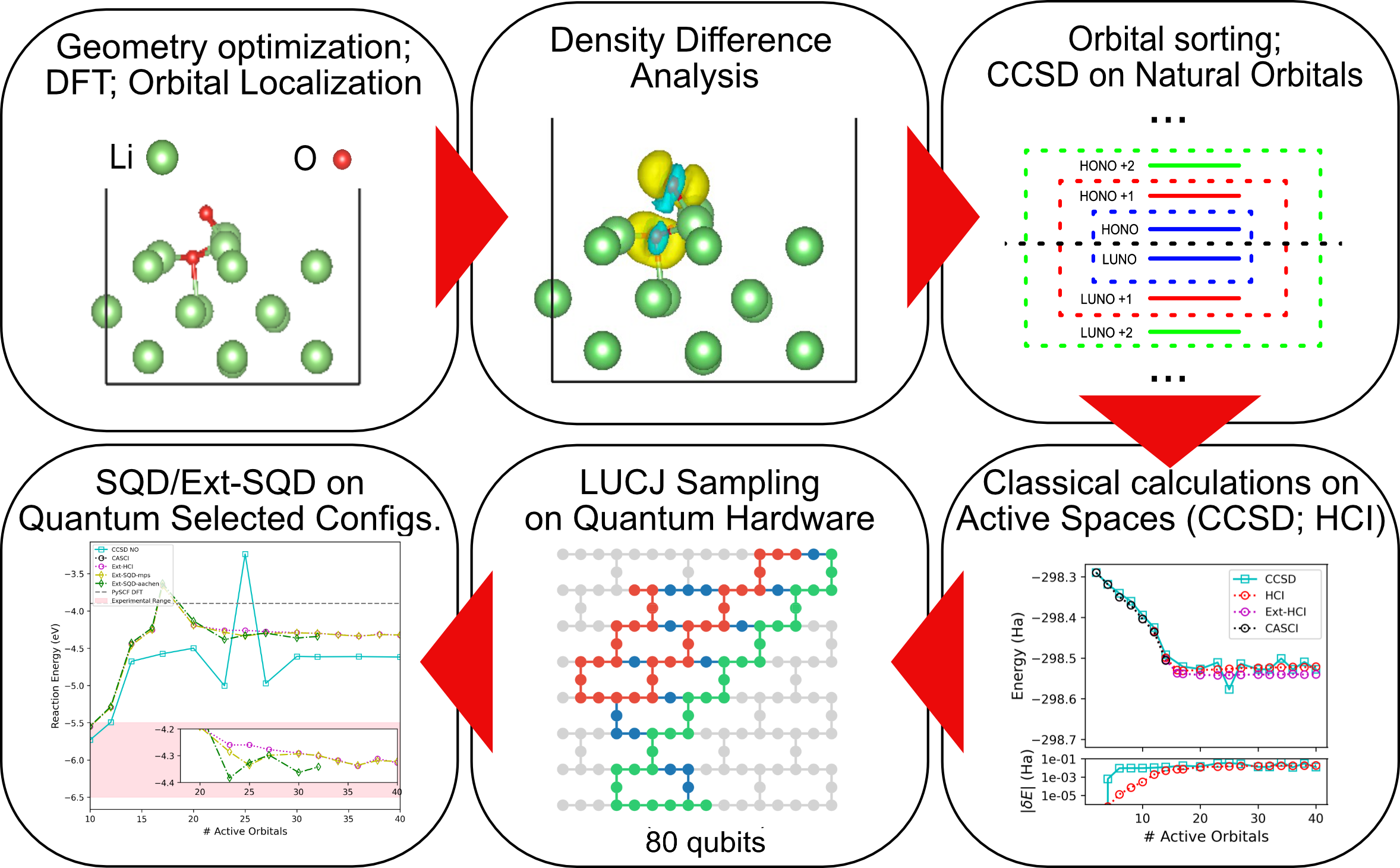}
\caption{\textbf{Hybrid, quantum-classical framework for studying lithium oxide formation in surface reactions.}  Geometry optimization and DFT calculations are followed by Active space selection via DDA for Natural Orbitals, and a LUCJ ansatz samples configurations from quantum hardware. Finally, SQD/Ext-SQD algorithms are used to compute the reaction energies $\Delta E = E_{\text{prod}} - E_{\text{reac}}$.} \label{fig:AS_worflow}
\end{figure}

To study the process, we have designed a computational workflow consisting of geometry optimization, active space selection, and quantum experiments. The workflow, summarized in \autoref{fig:AS_worflow}, enables a systematic investigation of lithium-oxygen interactions at the atomic level. The reaction energy is computed as the energy difference $\Delta E = E_{\text{prod}} - E_{\text{reac}}$ between reactant and product states. Predicting the reaction energy allows kinetic  analysis of lithium oxide formation, which is essential for battery stability.

\subsection*{Geometry Optimization}

% \begin{figure}[!htb]
% \centering
% \includegraphics[width=0.8\textwidth]{imgs/workflow_2.png}
% \caption{\textbf{Hybrid quantum-classical framework for studying lithium–oxygen reactions.}  Geometry optimization and DFT calculations lead into the Active space selection via DDA and Natural Orbitals which identifies the key orbitals, and a LUCJ ansatz samples configurations from quantum hardware. SQD/Ext-SQD then compute reaction energies ($\Delta E = E_{\text{prod}} - E_{\text{reac}}$), providing insights into lithium oxide formation.} \label{fig:AS_worflow}
% \end{figure}
All geometry optimizations are performed using DFT as implemented in the Vienna Ab initio Simulation Package (VASP) \cite{Kresse1993, Kresse1996, Kresse1999}. The structures are treated as periodic in the X-Y plane, with an Li slab thickness of 6.1 \AA{} and a vacuum layer of 12 \AA{} along the Z direction to minimize interactions between periodic images. The Brillouin zone is sampled using a $4 \times 4 \times 1$ \textbf{k}-point grid within the Monkhorst-Pack scheme. The oxygen molecule is allowed to relax on the surface of Li slab to ensure an accurate representation of surface interactions. The inert core electrons are accounted for by PAW pseudopotentials, and valence electrons are represented by plane-wave basis set. Only forces and geometries are allowed to relax while cell volume and shape are fixed. The convergence tolerance is set to $10^{-6}~\text{eV}$. 

In a next step, we utilize the GTH-DZV basis set \cite{GTH}, in combination with the Perdew-Burke-Ernzerhof (PBE) exchange-correlation functional \cite{PBE} and Goedecker-Teter-Hutter (GTH) pseudopotentials \cite{GTH}, to calculate the energy values and electronic densities for both reactant and product states using DFT within PySCF. \cite{pyscf} The electronic structure calculations are restricted to the $\Gamma$-point of the supercell as including additional \textbf{k}-points does not significantly alter the electronic density.

We find that the predicted reaction energy of Lithium oxide at the $\Gamma$-point, $\Delta E = -3.897~\text{eV}$, is smaller than the literature values at about $-6~\text{eV}$. \cite{ThermoTables,Radin2012, Hummelshoj2010-pv, Seriani2009-mn}

\subsection*{Active Space Selection}
% For many chemical systems, accurate results can be obtained by restricting electron correlation to a carefully selected subset of electrons and orbitals within an active space. Typically, the most effective choice involves valence electrons and orbitals, with further reductions permitted when chemically justified. It is crucial, however, that all orbitals contributing significantly to static correlation are included to maintain accuracy. \cite{Keller2015-zw, Stein2016-sz, Sayfutyarova2017-rl} 

% Studies have demonstrated that, in certain cases involving molecular adsorbates on surfaces, electron correlation effects are largely confined to a small number of orbitals and electrons, which suggests that constructing compact active spaces for surface reactions is feasible.\cite{Hirjibehedin2007-ud,Rau2014-am,Baumann2015-eo,Albertini2015-uo} To construct such an active space we employ methods as described in our previous work.\cite{Gujarati_2023}

 The active space selection process combines multiple techniques to systematically identify the most relevant electronic orbitals for quantum simulations. First, Kohn-Sham orbitals are localized using the Pipek-Mezey method to preserve electronic structure characteristics across the Brillouin zone. The localization isolates occupied and virtual orbitals that contribute to the reaction under study. Next, DDA is employed to assess charge redistribution by computing the electronic density difference between the reactant and product states. A tunable threshold function is then applied to filter out orbitals with negligible contributions, ensuring a physically meaningful selection of orbitals for the active space. Seven occupied orbitals are chosen for the reactant and the product as well as all virtual orbitals. The selections are based on the valence space of \ce{O2} and an initial evaluation of the computed electronic density for the orbitals.

To enhance convergence and accuracy, the active space is further refined using natural orbitals derived from coupled-cluster singles and doubles (CCSD) calculations performed in the initial active space. The selected orbitals are ranked based on occupation numbers, distinguishing those that are fully occupied, unoccupied, or fractionally occupied. In our systems, fractionally occupied orbitals do not occur. The refined active space is then constructed around the Highest Occupied Natural Orbital, HONO, and the Lowest Unoccupied Natural Orbital, LUNO, with additional orbitals included in an ``inside-out'' fashion for balancing accuracy and computational efficiency as needed.

\begin{figure}[!htb]
    \centering
    \begin{subfigure}[t]{0.45\textwidth}  % First figure
        \centering
        \includegraphics[width=\textwidth]{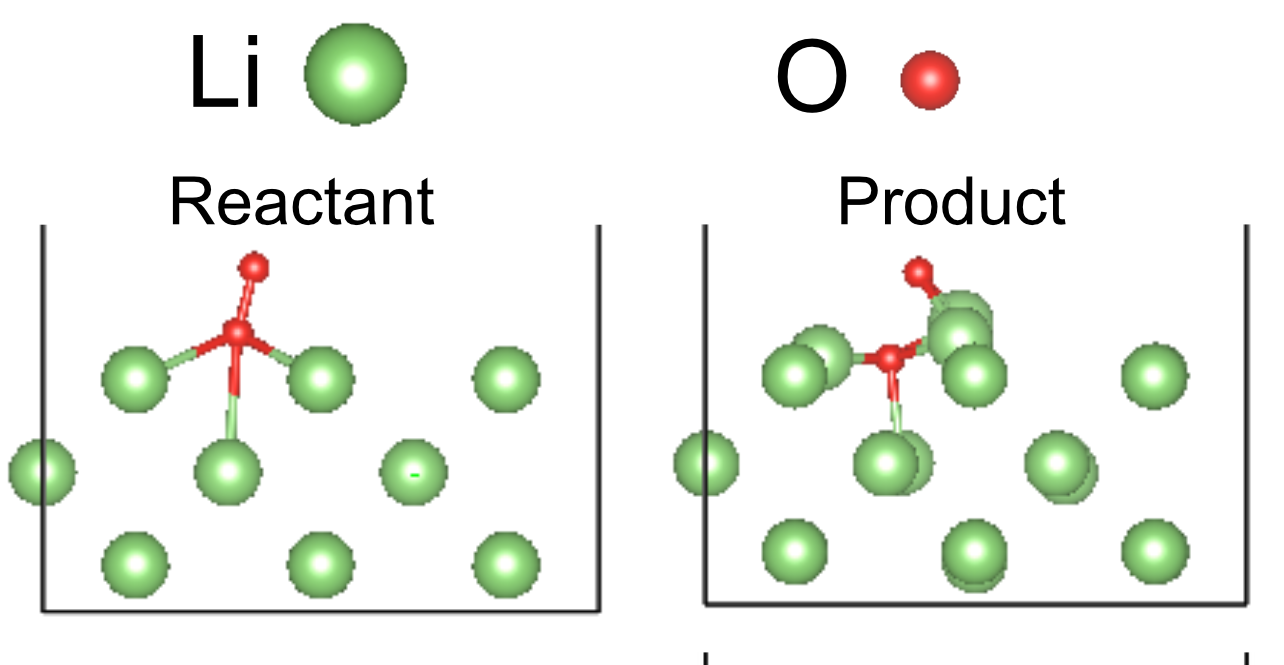}
        \caption{}
        \label{fig:geom_opt}
    \end{subfigure}
    \hfill
    \begin{subfigure}[t]{0.45\textwidth}  % Second figure
        \centering
        \includegraphics[width=\textwidth]{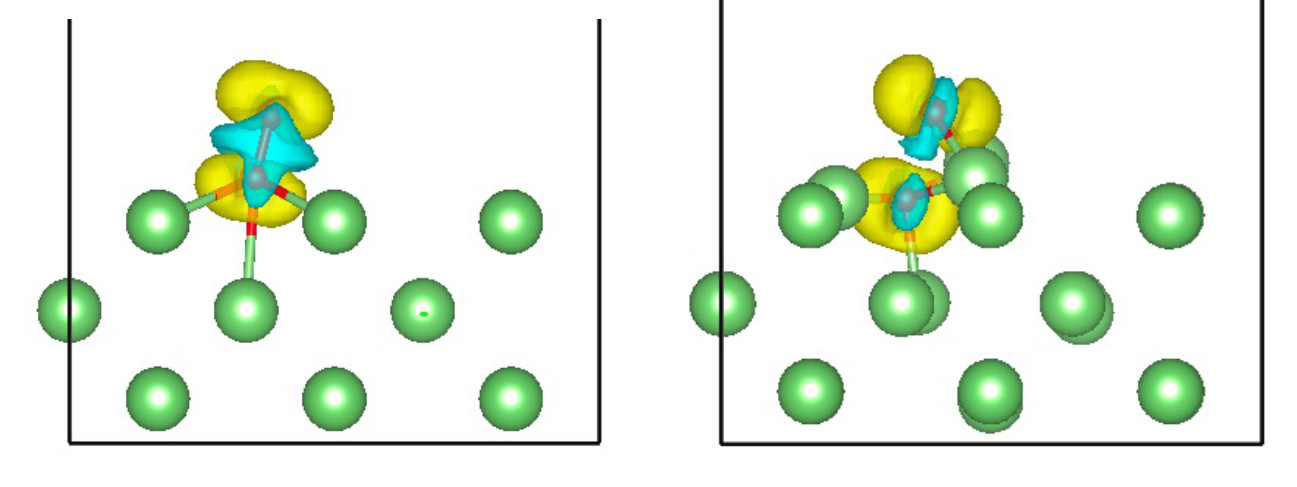}
        \caption{}
        \label{fig:DD}
    \end{subfigure}
    
    \vspace{1em}  % Adds vertical space before the next row

    \begin{subfigure}[t]{0.99\textwidth}  % Third figure centered below
        \centering
        \includegraphics[width=\textwidth]{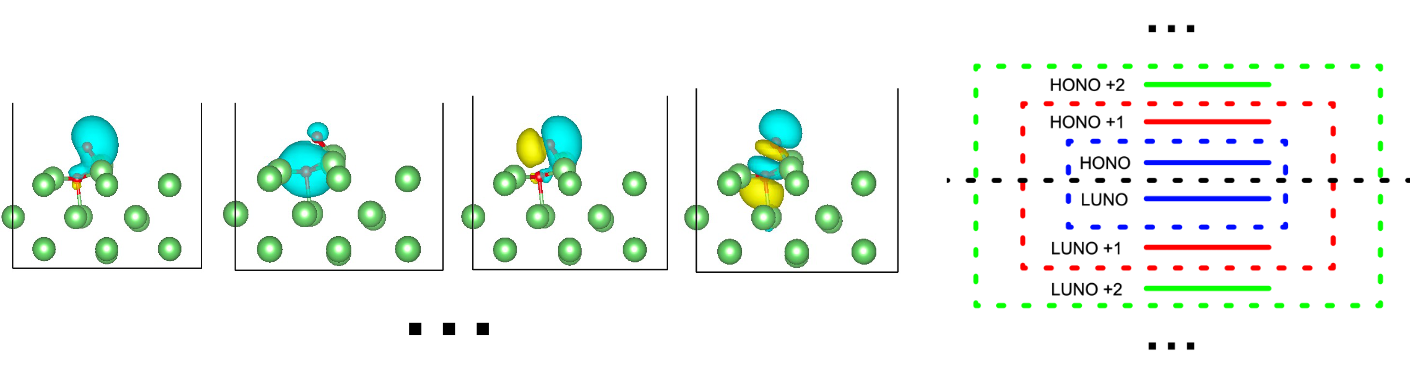}
        \caption{}
        \label{fig:Active_Space}
    \end{subfigure}

    \caption{\textbf{Active Space selection} Geometry optimization step (a). Density Difference Analysis step (b). Evaluation of each orbital's contribution to the density difference. Construction of active space using the Natural Orbitals (c), where orbitals for the product are shown. Orbitals are sorted based on their contribution to the density difference. The active space grows from the HONO-LUNO level ``inside out''.}
    \label{fig:AS_results}
\end{figure}

In \autoref{fig:AS_results}, we visualize the Active Space Selection process and the respective electronic densities for the system under investigation.

% \subsection*{Energy estimation}

% Once the active-space orbitals were determined for the $\Gamma$-point, the rest were fixed and frozen at their Hartree-Fock (HF) levels, and the Born-Oppenheimer Hamiltonian was systematically reduced to the active space using established methods. The Hamiltonian was then reformulated in second quantization and converted into a qubit representation through standard fermion-to-qubit transformations. For this, we utilized the Jordan-Wigner (JW) encoding.

\subsection*{Quantum-chemical Simulations}

Once the active-space orbitals are determined at the $\Gamma$-point, the remaining ones are fixed and frozen at their Hartree-Fock (HF) levels, and the Born-Oppenheimer Hamiltonian is systematically reduced to the active space applying standard methods. We use the CCSD calculations of the active-space Hamiltonian, as well as Complete Active Space Configuration Interaction (CASCI) and Heat-bath Configuration Interaction (HCI) calculations, for benchmarking the quantum algorithms. \cite{Boguslawski_2011, Holmes2016-jk}

For further refining the HCI results, we apply excitation operators to the final ground state following a method referred to as Extended Heat-bath Configuration Interaction, or Ext-HCI. The selection of higher-order excitations improves accuracy, albeit at an increased computational expense. We discard configurations in the wavefunction with amplitudes below a threshold of $10^{-2}$. Furthermore, we apply double excitation operators to configurations with amplitudes above $10^{-1}$, while applying single excitations to all others.

% \begin{figure}[h]
% \centering
% \includegraphics[width=0.9\textwidth]{imgs/classical_sims.png}
% \caption{Classical simulations for a given active space size. The top curves are for the reactant and the bottom ones are for the product. Energy values start to stabilize at 20 orbitals and remain so up until 40 orbitals. CASCI remains the most accurate value and is what other methods are benchmarked against, unfortunately scaling it beyond 14 orbitals is a challenge.}\label{fig:AS_classical}
% \end{figure}

\begin{figure}[!htb]
    \centering  % Aligns at top
    \includegraphics[width=\textwidth]{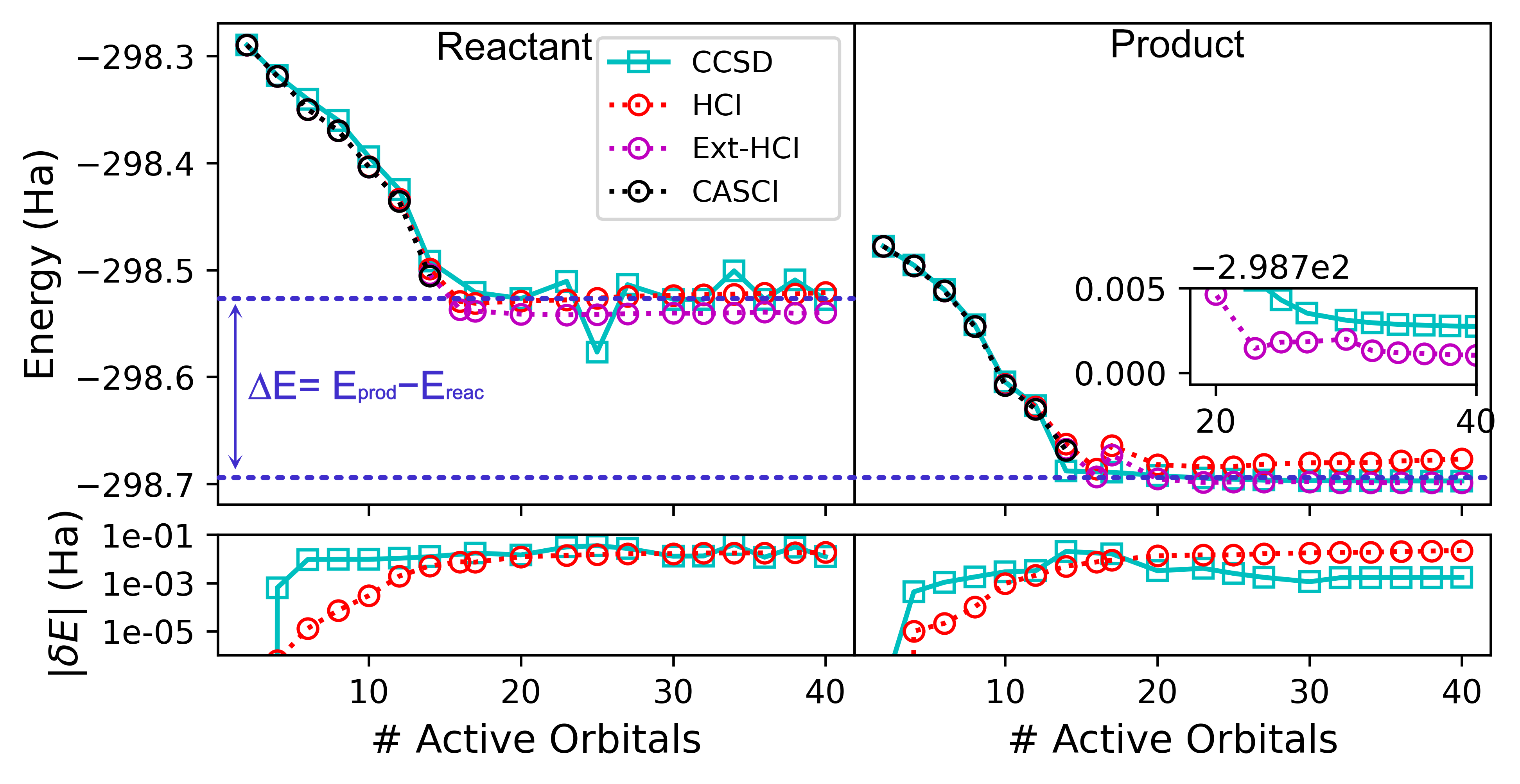}
    \caption{\textbf{Ground state and surface reaction energies obtained with standard, quantum-chemical simulation methods.} Ground state energies as a function of active space size. The curves on the left-hand side represent the reactant, the curves on the right-hand side the product. Energy reaction can be obtained by calculating the difference.}
    \label{fig:SQD_classical}
\end{figure}

In \autoref{fig:SQD_classical}, we show the results obtained with the "classical", quantum-chemical reference methods. Energy values converge at around 20 orbitals and continue to decrease up to 40 orbitals for HCI, Ext-HCI and CCSD. The results suggest that an active space of 40 orbitals would be necessary to calculate the energy difference. However, due to resource limitations, we are not able to scale CASCI calculations beyond 14 orbitals, but we can see reasonable agreement between the other three methods with errors within a few tens of mHa in this range. Therefore, Ext-HCI becomes the standard reference benchmark for higher orbital counts. For this reason, errors are calculated with respect to Ext-HCI, the most accurate, quantum-chemical reference method used in our investigation. As expected, the prediction error increases as we add orbitals to the active space.

\subsection*{Quantum Sampling Algorithms}
For the quantum experiments, we reformulate the Hamiltonian in second quantization and convert it into a qubit representation by means of fermion-to-qubit transformations using Jordan-Wigner (JW) encoding. SQD then projects the electronic Hamiltonian onto a subspace of sampled electronic configurations, ensuring proper electron count and spin sector. Similar to Quantum Selected Configuration Interaction (QSCI)~\cite{kanno2023quantumselectedconfigurationinteractionclassical}, SQD systematically expands the subspace for controlled accuracy improvements while maintaining computational efficiency. Our calculations involved $5-10$ iterations, with $16$ batches per iteration.

\begin{figure}[!htb]
\centering
\includegraphics[width=0.9\textwidth]{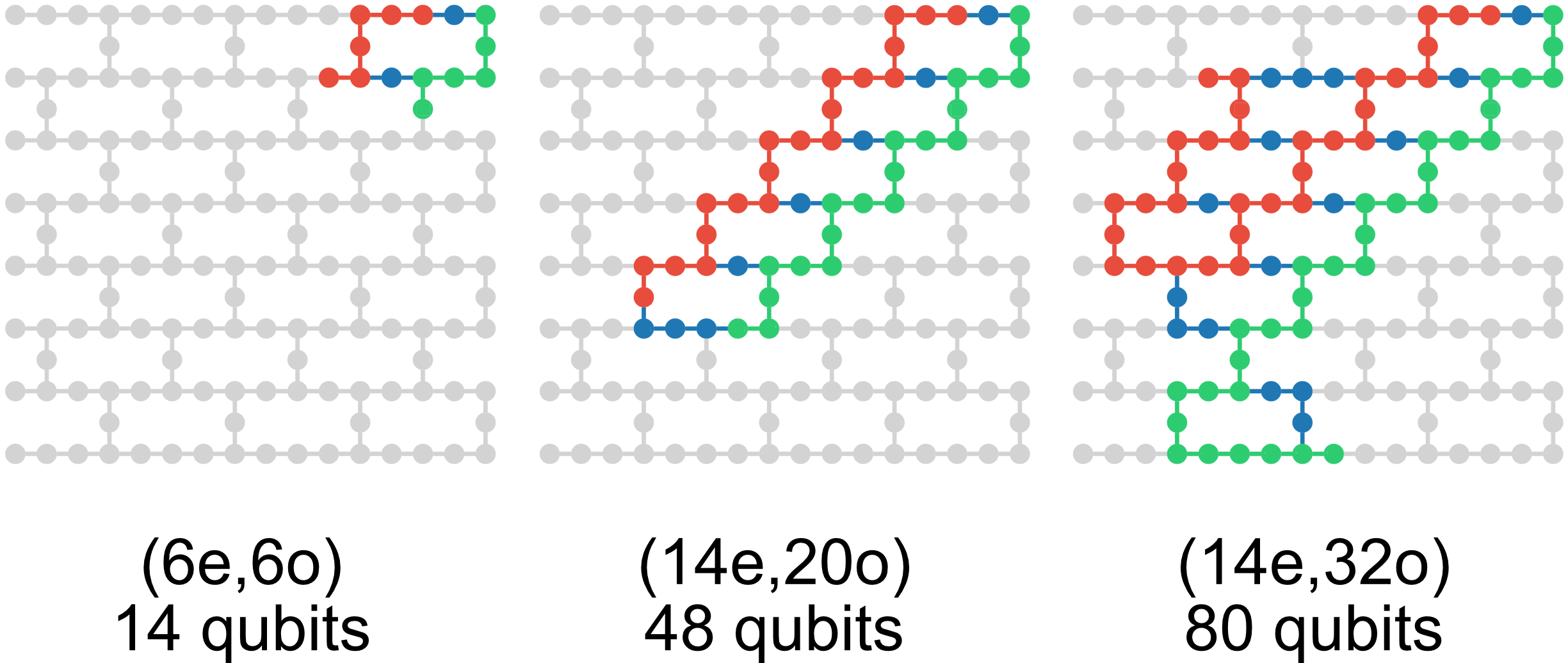}
\caption{\textbf{Circuit-to-chip mapping.} Green qubits represent $\alpha$ spin orbitals, red qubits represent $\beta$ spin orbitals, and blue qubits act as ancillae.}\label{fig:circ_diagrams}
\end{figure}

To enhance sampling quality, we use a truncated Local Unitary Cluster Jastrow (LUCJ) ansatz~\cite{Motta_2023}, omitting final orbital rotations and Jastrow interactions to balance accuracy and efficiency. As shown in \autoref{fig:circ_diagrams}, we initialize circuit parameters on the one- and two-body amplitudes taken from the CCSD simulations. We build quantum circuits using the \texttt{ffsim}\cite{ffsim} package and execute them on the \texttt{ibm\_aachen} device, optimizing qubit selection based on calibration data using \texttt{mapomatic}\cite{mapomatic}, thus excluding qubits with high error rates. Each circuit is sampled $6 \times 10^6$ times, and results are processed as bitstring distributions for SQD post-processing.

For improving sampling quality in large circuits representing 20 orbitals or more, we include additional ancillae after transpilation of the circuit with \texttt{mapomatic}. %This is specially necessary above 20 orbitals to improve sampling quality.

\begin{figure}[!htb]
    \centering
    \begin{subfigure}[t]{0.99\textwidth}  % Aligns at top
        \centering\includegraphics[width=\textwidth, trim=0 7 0 7, clip]{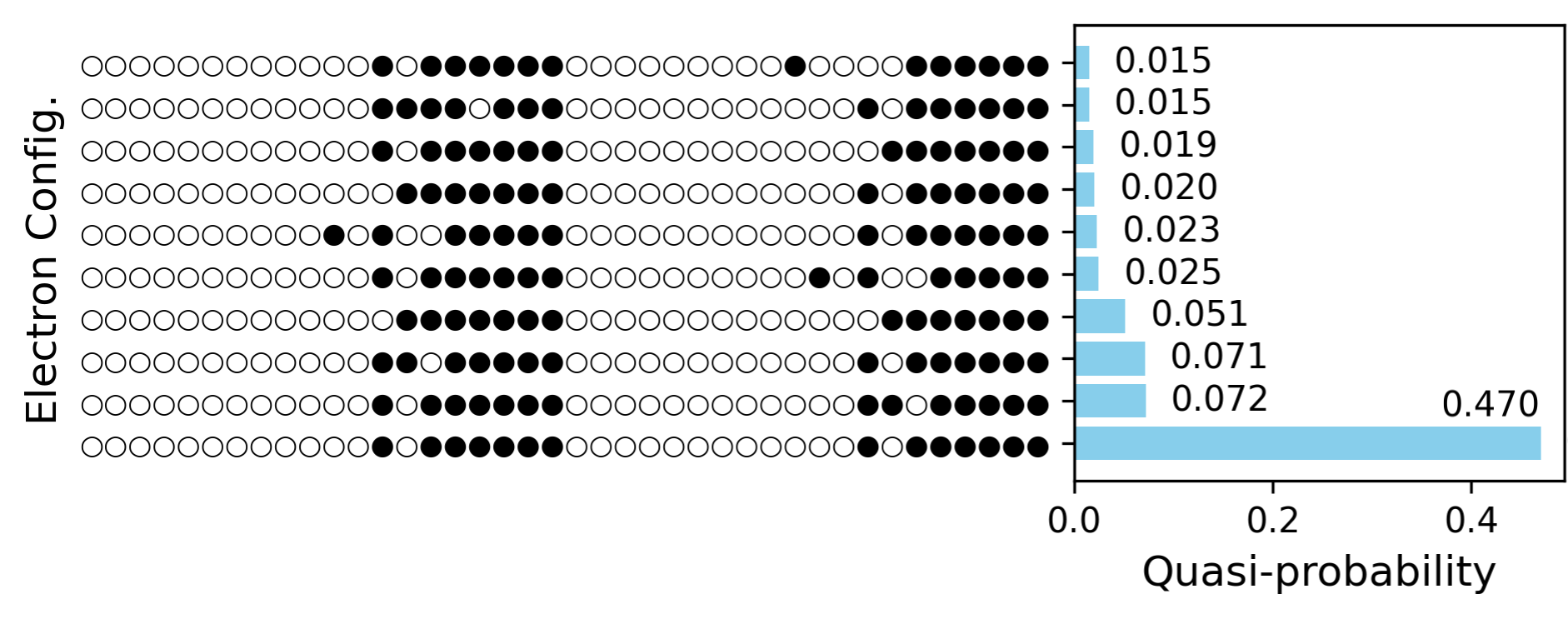}
        \caption{}
        \label{fig:histogram_HF}
    \end{subfigure}
    \begin{subfigure}[t]{0.99\textwidth}  % Aligns at top
        \centering\includegraphics[width=\textwidth, trim=0 10 0 10, clip]{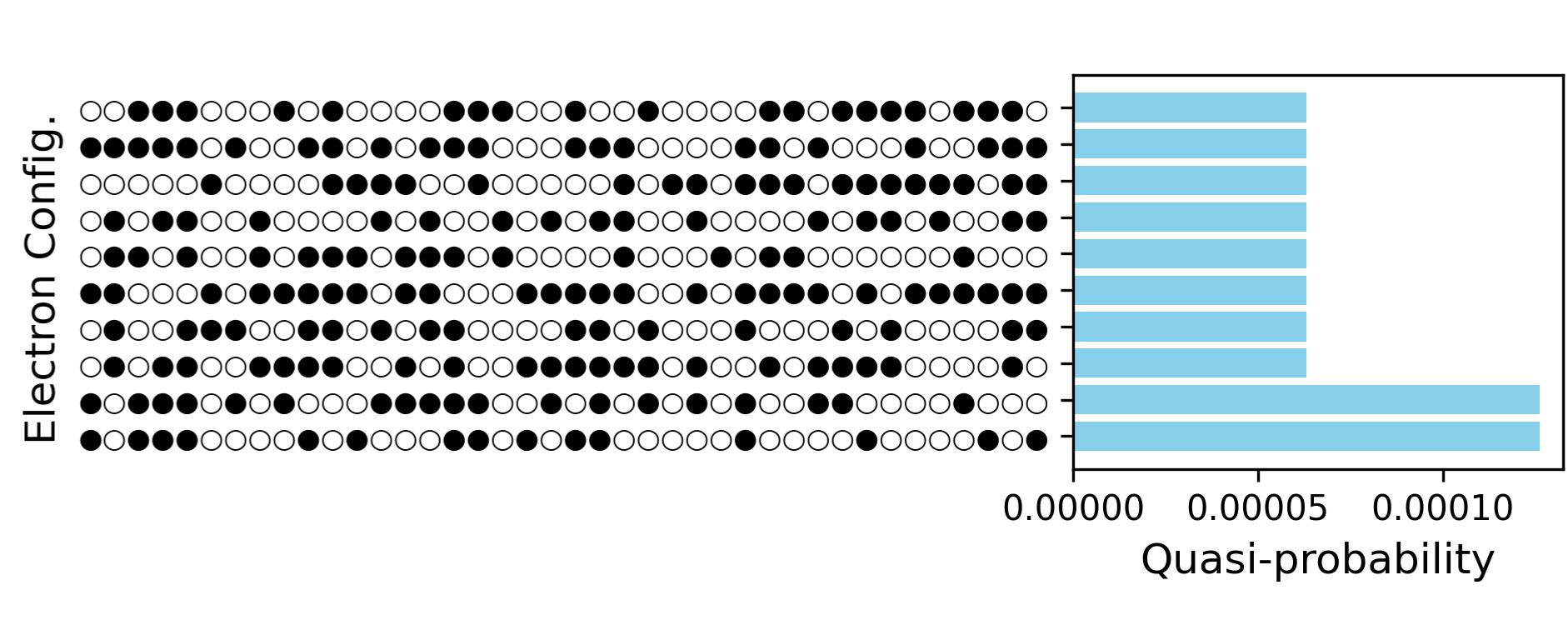}
        \caption{}\label{fig:histogram_20o}
    \end{subfigure}
    \caption{\textbf{Histograms of the top-10 electron configurations sampled by the quantum processing unit for the (14e,20o) reactant active space.} Filled (unfilled) circles represent occupied (unoccupied) orbitals. (a) Histogram for the non-truncated, one-layer LUCJ ansatz. (b) Histogram for the truncated, two-layer LUCJ ansatz. The truncated variant exhibits a distribution of electronic configurations that is not dominated by the reference Hartree-Fock state.}
    \label{fig:histogram}
\end{figure}

% Circuits are sampled on the \texttt{ibm\_aachen} device utilizing its heavy-hex topology in such a way that favors electronic structure calculations according to \autoref{fig:circ_diagrams}.\cite{Motta_2023} When mapping circuits to the chip we also take into account calibration data in order to avoid qubits with higher error rates. Each circuit was sampled around $10^6$ times, and results were collected as a counts dictionary with bitstrings representing configurations, which are then passed to the SQD calculation. 

In \autoref{fig:histogram}, we show histograms representing the output of the LUCJ circuit samples for the reactant in  a (14e, 20o) active space, that is, with 14 electrons distributed in 20 orbitals. In \autoref{fig:histogram_HF}, Hartree-Fock and its excitations are the most common configurations obtained with a non-truncated, one-layer LUCJ ansatz~\cite{robledomoreno2024chemistryexactsolutionsquantumcentric}. In \autoref{fig:histogram_20o}, we plot the results obtained with the truncated, two-layer circuit for comparison. Here, the distribution of configurations is not dominated by the HF state. For the purpose of our scaling study, therefore, we use the truncated two-layer LUCJ ansatz.

% \begin{figure}[h]
% \centering
% \includegraphics[width=0.9\textwidth]{imgs/histogram_23o.png}
% \caption{\textbf{Histograms representing the output from quantum hardware for the reactant in the active space of (14e,23o).} We display only the top 10 configurations, filled balls represent occupied orbitals, and unfilled ones are unoccupied orbitals.}\label{fig:histogram}
% \end{figure}

\begin{table}[htb]
\caption{\textbf{Metrics of representative quantum circuits.} $q$ is the total number of qubits, which is the sum of $s$ qubits representing spin orbitals and $a$ ancilla qubits. $d$ is the 2-qubit circuit depth, $CX$ is the number of CNOT gates and $u$ the number of 1-qubit gates. } 
\label{tab:quantum_resources}
\centering
\begin{tabularx}{\textwidth}{@{}lXXXXX@{}}
\toprule
\textbf{System} & $q (s+a)$ & $d$ & $CX$ & $u$ \\
\midrule
Prod(6e,6o)     &14(12+2)&  74  & 132  & 224 \\
Prod(12e,12o)   &30(24+6)&  124  & 474  & 872   \\
Prod(14e,20o)   &48(40+8)&  188 & 1228  & 2345  \\
Prod(14e,32o)   &80(64+16)&  466 & 3026  & 5551  \\
\midrule
Reac(6e,6o)     &14(12+2)&  72  & 132  & 222 \\
Reac(12e,12o)   &30(24+6)&  124  & 474  & 870 \\
Reac(14e,20o)   &48(40+8)&  184 & 1194  & 2289   \\
Reac(14e,32o)   &80(64+16)&  392 & 2874  & 5388  \\
\bottomrule
\end{tabularx}
\end{table}

In \autoref{tab:quantum_resources}, we show circuit sizes and qubit requirements for representative circuits. The circuit size is determined by the depth and number of gates. Our circuit requirements are similar to those reported in reference \cite{robledomoreno2024chemistryexactsolutionsquantumcentric}. Bitstrings in the distribution sampled by \texttt{ibm\_aachen} have on average 93.57\% of forbidden configurations, i.e., incorrect Hamming weights due to noise. This is around 10 times better than a uniform configuration distribution with an average 99.922\% of forbidden configurations. At about 32 orbitals, the circuits become considerably more complicated. With 2935 CNOT gates, we are close to the practical noise limit. Here, the average is at 96.54\% for the reactant and at 96.28\% for the product, which negatively affects the calculated energy values for SQD. Consequently, the accuracy of the results continues to deteriorate towards higher orbital counts. Details are available in the \autoref{sec:SQD_results} section of the Supplementary Information. To further enhance the accuracy of molecular ground-state energies beyond SQD, we will explore in the following the Ext-SQD method.

Ext-SQD increases accuracy by applying excitation operators to sampled configurations~\cite{barison2024quantumcentriccomputationmolecularexcited}. For comparison, we discard configurations in the wavefunction with amplitudes below $10^{-2}$ and apply double excitation operators to configurations that have amplitudes larger than $10^{-1}$, while including single excitations for the remainder. Ext-SQD calculations do not rely on configuration recovery and are based on the results of the preceding SQD execution, so only one iteration is necessary. We keep the number of batches at $16$.

\begin{figure}[!htb]
    \centering
    \begin{subfigure}[t]{0.99\textwidth}  % Aligns at top
        \centering\includegraphics[width=\textwidth]{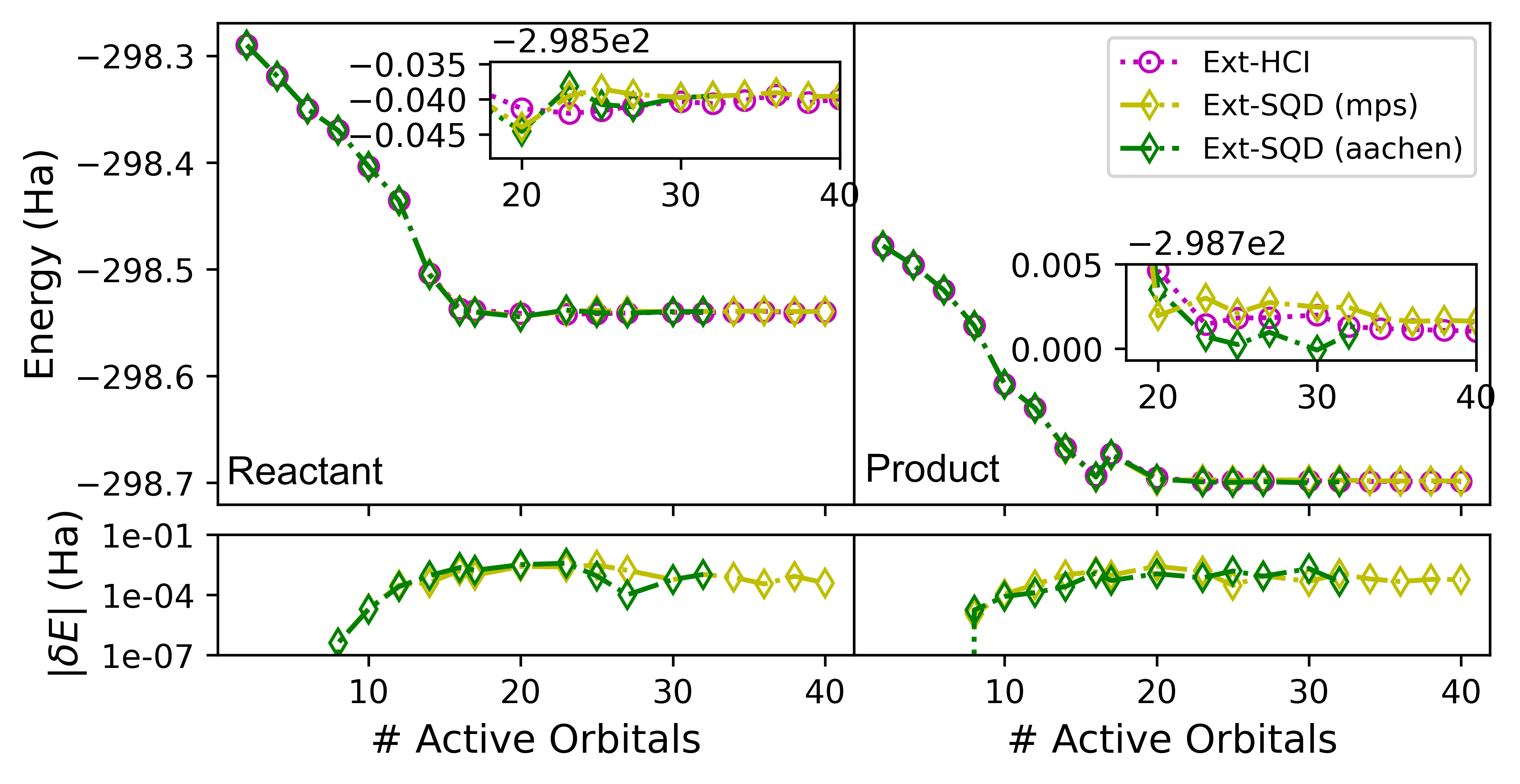}
        \caption{}
        \label{fig:SQD_quantum}
    \end{subfigure}
    \vspace{1em}  % Adds vertical space before the next row
    \begin{subfigure}[t]{0.99\textwidth}  % Aligns at top
        \centering
        \includegraphics[width=\textwidth]{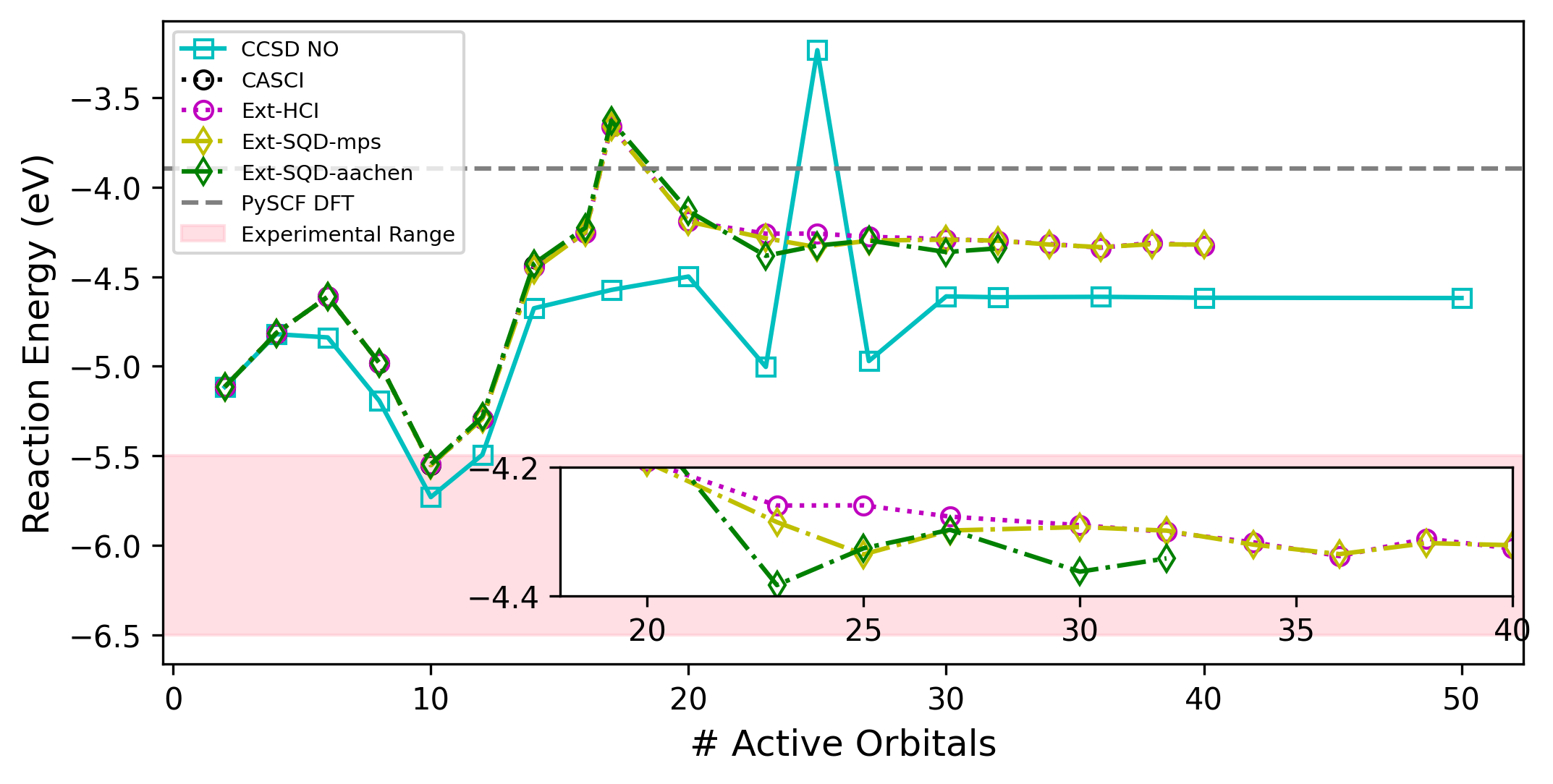}
        \caption{}
        \label{fig:Energy_delta}
    \end{subfigure}
    \caption{\textbf{Comparison of ground state and reaction energies obtained with the quantum processing unit} (a) Simulated (yellow diamonds) and experimental (green diamonds) Ext-SQD results compared to Ext-HCI, the classical reference standard. Ext-SQD experiments can be performed for active space sizes up to 32 orbitals. (b) Reaction energies computed at the $\Gamma$-point with quantum and classical methods, respectively.  DFT results are shown as a reference.}
    \label{fig:SQD_combined}
\end{figure}

As shown in \autoref{fig:SQD_quantum}, the Ext-SQD energy predictions achieve higher accuracy than SQD, consistently close to those produced by Ext-HCI. As compared to both HCI and CCSD, Ext-SQD results satisfy the $1.6~\text{mHa}$ chemical accuracy bound. We observe that Ext-SQD reaches lower energy values for both product and reactant above 14 orbitals. Scaling Ext-SQD beyond 32 orbitals is challenging, as the quality of quantum-sampled configurations deteriorates. At and above 34 orbitals, we are unable to obtain configurations with correct Hamming weights within a set of $6 \times 10^6$ samples.

At this scale, as shown in \autoref{fig:circ_diagrams}, the processor topology limits the mapping of circuits to the lattice diagonals. Adding orbitals makes it increasingly difficult to design circuits that avoid certain qubits and gates with high error rates, reducing sample quality overall.

For scaling SQD and Ext-SQD experiments up to 40 orbitals, we approximate the LUCJ quantum circuits with tensor networks using a truncation value of $10^{-8}$ and Matrix Product State (MPS) simulators. \cite{Robertson_2025,qiskit-addon-aqc-tensor} While we are indeed able to obtain the results, this approach produces slightly higher energy values, due to the truncation step, as can be seen in \autoref{fig:SQD_quantum}.

The post-processing calculations are conducted on HPC nodes with 192 x86 cores and 3 TB of RAM by means of the Dice diagonalizer \cite{qiskit-addon-dice-solver}. As shown in \autoref{tab:SQD_resources}, as the number of orbitals increases, the subspace dimension for Ext-SQD grows faster than for standard SQD.  While the subspace dimension for Ext-SQD is consistently smaller than for Ext-HCI, it is still of the same order of magnitude. This is because the number of single excitations grows quadratically with the number of orbitals, while the number of double excitations grows with the fourth power. At higher orbital counts, this limits the number of double excitations that can be included in Ext-SQD~\cite{barison2024quantumcentriccomputationmolecularexcited}.

In \autoref{fig:Energy_delta}, we plot the reaction energy for active spaces with up to 50 orbitals and the DFT-values as a reference. Reaction energy values for representative active spaces are provided in \autoref{tab:En_delta}. HCI, Ext-SQD and CCSD provide lower energy values, closer to those found in literature, which usually rely on empirical corrections~\cite{Radin2012, Galloway2019-lv, Hummelshoj2010-pv, Seriani2009-mn}. While we are unable to consistently scale beyond 40 orbitals, the energy values produced with Ext-HCI and Ext-SQD exhibit convergence at smaller active space sizes than in the case of CCSD. While CCSD provides a $\Delta E$-value closer to experimental references, it overestimates the reactant energy. We note that CCSD, unlike CI based methods such as CASCI, HCI, Ext-HCI, SQD and Ext-SQD\cite{robledomoreno2024chemistryexactsolutionsquantumcentric,kanno2023quantumselectedconfigurationinteractionclassical},  is not a variational method. This means it can predict values below the actual ground state energy \cite{Cremer2013-ua, olsen_bible}, which complicates the comparison.

Using atomic coordinates derived from DFT relaxations, all subsequent prediction methods are expected to be affected by the mean-field nature of DFT in defining the reactant and product states. Nevertheless, the prediction results improve upon the initial reaction energy values overall, successively approaching the experimental range.

\begin{table}[htb]
\caption{\textbf{Ext-SQD and Ext-HCI calculations for product and reactant in representative active spaces.} $|\tilde{\chi}|$ is the number of samples taken from \textbf{ibm\_aachen}, $K$ is the number of configuration recovery iterations performed in SQD, $D$ is the dimension of the subspace diagonalized by SQD, $D_E$ is the dimension of the subspace diagonalized by Ext-SQD, and $D_{EH}$ is the dimension of the subspace diagonalized by Ext-HCI.} 
\label{tab:SQD_resources}
\centering
\begin{tabularx}{\textwidth}{@{}lXXXXX@{}}
\toprule
\textbf{System} & \boldmath{$|\tilde{\chi}|$} & \boldmath{$K$} & \boldmath{$D$} & \boldmath{$D_{E}$} & \boldmath{$D_{EH}$}\\
\midrule
Prod(2e,2o)     & $1.5\cdot10^{6}$  & 5  & $2.7\cdot10^{4}$  & $1.5\cdot10^{4}$ & $1.5\cdot10^{4}$ \\
Prod(12e,12o)   & $1.5\cdot10^{6}$  & 7  & $3.4\cdot10^{5}$  & $2\cdot10^{5}$ & $5.3\cdot10^{5}$  \\
Prod(14e,20o)   & $2\cdot10^{6}$    & 10 & $1.4\cdot10^{6}$  & $7.9\cdot10^{6}$ & $9.6\cdot10^{6}$ \\
Prod(14e,30o)   & $2\cdot10^{6}$    & 10 & $1.4\cdot10^{6}$  & $33\cdot10^{6}$ & $84\cdot10^{6}$ \\
\midrule
Reac(2e,2o)     & $1.5\cdot10^{6}$  & 5  & $2.9\cdot10^{4}$  & $1.2\cdot10^{4}$ & $1.2\cdot10^{4}$ \\
Reac(12e,12o)   & $1.5\cdot10^{6}$  & 7  & $3.4\cdot10^{5}$  & $1.8\cdot10^{5}$ & $4.7\cdot10^{5}$\\
Reac(14e,20o)   & $2\cdot10^{6}$    & 10 & $1.3\cdot10^{6}$  & $9\cdot10^{6}$  & $11\cdot10^{6}$ \\
Reac(14e,30o)   & $2\cdot10^{6}$    & 10 & $1.3\cdot10^{6}$  & $28\cdot10^{6}$ & $56\cdot10^{6}$ \\
\bottomrule
\end{tabularx}
\end{table}

\begin{table}[htb]
\caption{\textbf{Comparison of reaction energies obtained in representative active spaces.} The $\Delta E$-values are provided in eV.} 
\label{tab:En_delta}
\centering
\begin{tabularx}{\textwidth}{@{}lXXXXX@{}}
\toprule
\textbf{Act. Space} & \textbf{CASCI} & \textbf{CCSD} & \textbf{Ext-HCI} & \textbf{Ext-SQD (MPS)} & \textbf{Ext-SQD (aachen)}\\
\midrule
(2e,2o)     & $-5.116$  & $-5.116$  & $-5.116$ & $-5.116$  & $-5.116$\\
(12e,12o)   & $-5.294$  & $-5.494$  & $-5.294$ & $-5.294$  & $-5.283$\\
(14e,20o)   & N/A  & $-4.500$  & $-4.191$ & $-4.193$  & $-4.134$\\
(14e,30o)   & N/A  & $-4.610$  & $-4.290$ & $-4.293$  & $-4.363$\\
(14e,32o)   & N/A  & $-4.615$  & $-4.301$ & $-4.299$  & $-4.342$\\
\bottomrule
\end{tabularx}
\end{table}

\section*{Discussion}\label{sec:Disc}

The results presented in this study demonstrate the potential of hybrid quantum-classical methods for improving the accuracy of electronic structure calculations. By leveraging a systematic, active-space selection approach and employing the SQD and Ext-SQD method, respectively, we are able to compute ground-state energies with higher accuracy than select standard reference methods, e.g., HCI and CCSD. The agreement of results obtained with Ext-SQD and CASCI for active spaces up to 12 orbitals, along with the scaling of Ext-SQD up to 32 orbitals, demonstrates the utility of the approach for studying electronic configurations in chemistry and materials science.

We note that the differences in energy values obtained with Ext-HCI are small enough to be considered chemically equivalent, i.e., $<1.6$mHa, while maintaining similar subspace dimensions, see \autoref{tab:SQD_resources}. The same applies to the calculated reaction energies.

The methodology outlined in \autoref{fig:AS_worflow}  integrates classical and quantum techniques to achieve computationally efficient simulations. The use of DDA and natural orbitals from coupled-cluster calculations ensures that the active space captures the relevant electronic degrees of freedom. The refinement of active spaces by means of coupled-cluster natural orbitals improves both convergence and accuracy of the quantum calculations.

The integration of quantum hardware in the computational architecture allowed us to validate the appraoch with a state-of-the-art quantum processor. As shown in \autoref{tab:quantum_resources}, circuit depths and gate counts increase significantly with the number of orbitals in the active space, limiting the upscaling of quantum simulations with current quantum processing units. The results obtained from executing LUCJ circuits confirm that our approach is feasible on current quantum devices, but noise mitigation strategies will be necessary to maintain accuracy at larger system sizes. Importantly, the simulation results indicate the feasibility of extended active spaces, beyond 32 orbitals, as soon as quantum processing units with lower error rates become available.

As a key finding of our investigation, Ext-SQD provides significant improvements over standard SQD by incorporating excitation operators to refine the sampled configuration space. The enhancement allows for better representation of ground-state wavefunctions while maintaining computational feasibility. As indicated in \autoref{fig:SQD_quantum}, however, the computational cost of Ext-SQD grows significantly as the number of orbitals increases, shifting the computational bottleneck from quantum to classical resources. This behaviour can bee seen between HCI and Ext-HCI at smaller scale, where both methods diverge for active spaces above 14 orbitals. Research and development is needed for optimizing classical post-processing techniques, such as parallelizing the diagonalization of extended subspaces. Such improvements could unlock scaling up to $40$ orbitals, where the computational bottleneck shifts back to the quantum stack.

An important result of our scaling study is the rigorous comparison with classical methods as function of active space size. Based on the analysis in \autoref{fig:SQD_combined}, we monitor how the energy differences computed with Ext-SQD converge earlier than those obtained with CCSD. Remarkably, at 20 orbitals, we observe that the Ext-SQD results surpass the energy values computed with both CCSD and HCI, marking the method's performance "sweet spot" given the current computational infrastructure.  This indicates that quantum methods, when properly integrated with classical techniques, can deliver improvements of both computational efficiency and accuracy with regard to classical reference methods.

So far, we have applied variational references (HCI/Ext-HCI, SQD/Ext-SQD) and CCSD in the active space. A natural question is the role of dynamic correlations outside of the active space. Established, multi-reference second-order perturbation theories, e.g., CASPT2 and NEVPT2, are designed to recover such correlation in CAS-quality~\cite{Andersson1992,Angeli2001sc}. In practice, CASPT2 often employs level-shift strategies, real or imaginary, and the IPEA shift to mitigate intruder states,\cite{Forsberg1997,Ghigo2004} while NEVPT2 is intruder-state free and rigorously size-consistent in its contracted formulations~\cite{Angeli2001sc}. 

From an algorithmic standpoint, our Ext-SQD wavefunctions provide multi-reference states within the active space for defining a subsequent PT2 corrections over external orbitals. This mirrors the SCI+PT strategy used with selected-CI references,\cite{Sharma2017,Li2018} and offers aroute to incorporate external-space, dynamic correlation without increasing quantum circuit depth. Although we do not report CASPT2/NEVPT2 results for our largest embedded systems, we note that a PT2 step lowers absolute energies but partially cancels out in reaction energies. The net effect depends on differential, external correlations between reactants and products. Implementing Ext-SQD with PT2 and NEVPT2 is, therefore, a plausible direction for scaling accuracy. We  outline the required steps in the Methods section.

Despite the advancements, several challenges must be addressed before quantum-assisted simulations become a practical tool for chemistry and materials applications. Noise resilience, circuit depth limitations, and efficient state preparation remain key obstacles. The LUCJ ansatz, while effective in reducing circuit depth, may require further modifications to improve expressivity and error resilience. Moreover, exploring alternative quantum ansatzes, such as adaptive variational circuits, could improve energy estimations while maintaining hardware efficiency.

Looking ahead, the insights gained from this work provide a strong foundation for future research in quantum-assisted simulations of surface reactions. The extension of our approach to more complex reaction networks, multi-step catalytic processes, and beyond lithium-oxygen systems will be an important direction. As quantum hardware continues to advance, the scalability of hybrid methods like Ext-SQD will be further tested, paving the way for practical applications in chemistry and materials science.

In conclusion, this study demonstrates that hybrid quantum-classical methodologies, particularly those leveraging advanced active space selection and sample-based diagonalization techniques, open a promising path for studying complex chemical problems. Our results underscore the potential of quantum computing to complement and extend classical approaches, bringing us closer to achieving accurate and efficient electronic structure calculations in materials discovery.

\section*{Methods}\label{sec:Meth}

\subsection*{DFT Simulations}

Following the procedure outlined in reference \cite{Gujarati_2023}, after performing a geometry optimization step, we conducted DFT calculations with a basis of translationally adapted, linear combinations of Gaussian atomic orbitals (AOs) which are expressed as:

\begin{equation}
\phi_{k,p}(r) = \sum_{T} e^{i k \cdot T} \chi_p (r - T),
\end{equation}

where \(T = \sum_{i=1}^{3} T_i a_i\) represents a lattice translation vector, \( k = \sum_{i=1}^{3} k_i b_i\) denotes a momentum vector within the first Brillouin zone and \( \chi_p \) is a Gaussian-type atomic orbital. The summation over \(T\) ensures that the constructed orbitals conform with the translational periodicity of the system\cite{McClain_2017,SolidStateChemistry_2005}.

All electronic structure calculations were performed using the PySCF package \cite{pyscf}, employing the GTH-DZV basis set \cite{GTH} and the Goedecker-Teter-Hutter (GTH) pseudopotentials, \cite{GTH} as well as the Perdew-Burke-Ernzerhof (PBE) exchange-correlation functional \cite{PBE}. 

Within this basis, the Born-Oppenheimer Hamiltonian takes the following representation:\cite{VANDEVONDELE2005103}

% \begin{equation}
% \hat{H}
% = E_0
% + \sum_{\substack{\mathbf{k}\\ p\,r\,\sigma}}
%   h_{pr}(\mathbf{k})\,\hat{a}^\dagger_{\mathbf{k} p \sigma}\hat{a}_{\mathbf{k} r \sigma}
% + \frac{1}{2}
%   \sum_{\substack{\mathbf{k}_p,\mathbf{k}_r,\mathbf{k}_q,\mathbf{k}_s\\ p\,r\,q\,s\,;\,\sigma\,\tau}}^{\!*}
%   (\mathbf{k}_p p,\mathbf{k}_r r \mid \mathbf{k}_q q,\mathbf{k}_s s)\,
%   \hat{a}^\dagger_{\mathbf{k}_p p \sigma}
%   \hat{a}^\dagger_{\mathbf{k}_q q \tau}
%   \hat{a}_{\mathbf{k}_s s \tau}
%   \hat{a}_{\mathbf{k}_r r \sigma}\,.
% \end{equation}

\begin{equation}
\begin{aligned}
\hat{H} \;=\; &\, E_0 
+ \sum_{\substack{\mathbf{k}\\ p\,r\,\sigma}}
  h_{pr}(\mathbf{k})\,
  \hat{a}^\dagger_{\mathbf{k} p \sigma}\,
  \hat{a}_{\mathbf{k} r \sigma}
\\[2pt]
&\;+\; \frac{1}{2}\,
   \sum_{\substack{\mathbf{k}_p,\mathbf{k}_r,\mathbf{k}_q,\mathbf{k}_s\\ p\,r\,q\,s\,;\,\sigma\,\tau}}^{\!*}
   \big(\,\mathbf{k}_p p,\;\mathbf{k}_r r \,\big|\, \mathbf{k}_q q,\;\mathbf{k}_s s\,\big)\,
   \hat{a}^\dagger_{\mathbf{k}_p p \sigma}\,
   \hat{a}^\dagger_{\mathbf{k}_q q \tau}\,
   \hat{a}_{\mathbf{k}_s s \tau}\,
   \hat{a}_{\mathbf{k}_r r \sigma}\,,
\end{aligned}
\end{equation}

where momentum conservation is enforced such that \( k_p + k_q - k_r - k_s = G \), with \( G \) being a reciprocal lattice vector. To efficiently approximate electron-electron interactions, we employed density fitting with a Weigend auxiliary basis.

\subsection*{Automated Active Space Selection}

We localized the Kohn-Sham orbitals using the Pipek-Mezey method \cite{Pipek_1989}, which relies on a Mulliken population analysis with meta-Löwdin orbitals \cite{Qiming_2014, Lehtola_2014}. Localization was performed separately for occupied and virtual orbitals at each individual \textbf{k}-point to ensure that the electronic structure characteristics were preserved across the Brillouin zone. We computed the electronic density difference at the DFT level, as defined by
\begin{equation}
\rho_{\text{DD}}(\mathbf{x}) = \rho_{\text{Li}+\text{O}_2}(\mathbf{x}) -\rho_{\text{O}_2}(\mathbf{x}) - \rho_{\text{Li}}(\mathbf{x}), 
\end{equation}
which quantifies the charge redistribution during the reaction, as shown in \autoref{fig:DD}. 

To automate the selection of active spaces, we ranked orbitals based on their contributions to the density difference. To quantify these contributions, we introduced a threshold parameter $\eta$ for suppressing minor orbital components and for retaining only physically meaningful orbitals. By systematically varying this threshold, we identified a range where orbital rankings remain consistent, facilitating a robust active space selection. Through a systematic scan, we identified a stability range of \( \eta \approx 10^{-3} \). Within this range, the ranking of occupied localized orbitals remained stable.

%This approach automates the selection of active spaces by focusing on the physics governing molecule-surface interactions, allowing for systematic improvements. The process begins with an initial mean-field calculation of the entire system to obtain a comprehensive electronic structure. From this, localized molecular orbitals are identified, which serve as the foundation for defining the active space. The active space is then treated with high-level quantum chemical methods, while the remainder of the system is described using lower-level methods.

To further enhance active space selection and ensure convergence with respect to active space size, we combined the density difference analysis with natural orbitals obtained from CCSD calculations. Specifically, an initial CCSD calculation was performed within an active space constructed from the highest-ranked occupied orbitals identified via the density difference analysis and all virtual orbitals. The resulting natural orbitals exhibit occupation numbers predominantly close to either zero or two, clearly distinguishing high-occupancy orbitals from low-occupancy ones. In cases involving stronger electron correlation, orbitals with fractional occupancy had to be explicitly included. The final active space was systematically constructed around the Highest Occupied Natural Orbital (HONO) and the Lowest Unoccupied Natural Orbital (LUNO). For further detail, we refer to reference \cite{Gujarati_2023}.

% To enhance convergence with respect to active-space size, we employed a method that combines density difference analysis and natural orbitals from coupled-cluster calculations (CCSD).

% In this approach, we first performed a CCSD calculation using an initial active space consisting of the highest-ranked occupied DFT orbitals (selected using the density difference method) along with all virtual orbitals. From this CCSD wavefunction, we extracted the natural orbitals as the eigenvectors of the one-particle density matrix and sorted them in descending order according to their occupation numbers. 

% For the systems studied, occupation numbers were predominantly close to either 2 (fully occupied) or 0 (unoccupied), which allowed for a clear distinction between high- and low-occupancy orbitals. In systems with strong electron correlation, where some orbitals have fractional occupation numbers, those orbitals must be explicitly included in the active space.

% We define the Highest Occupied Natural Orbital ($\text{HONO} = N_e/2$) and the Lowest Unoccupied Natural Orbital ($\text{LUNO} = N_e/2 + 1$), where \( N_e \) represents the number of electrons in the system. The final active space was constructed by selecting orbitals in the range from $\text{HONO} - \min(k, N_e/2 - 1)$ to $\text{LUNO} + \min(k, N_o - N_e/2 - 1)$, where \( k \) is a positive integer, and \( N_o \) is the total number of orbitals.

\subsection*{Sample-based Quantum Diagonalization }

 The Sample-based Quantum Diagonalization (SQD)~\cite{robledomoreno2024chemistryexactsolutionsquantumcentric, Nutzel_2025,kanno2023quantumselectedconfigurationinteractionclassical} approach involved projecting the electronic Hamiltonian onto a subspace defined by a collection of electronic configurations sampled from a probability distribution. These set of configurations, denoted as $X = \{x\}$, were drawn from the wavefunction probability \(P_{\Psi}(x) = |\langle x | \Psi \rangle |^2\), whereas a noisy version is represented by \(P_{\tilde{\Psi}}(x) = \langle x | \tilde{\rho} | x \rangle\). The subset of configurations that maintain the correct particle number was labeled $X_N$, and through a self-consistent recovery process an expanded configuration set $X_R$ was constructed. The electronic Hamiltonian $\hat{H}$ was then projected onto the selected subspace, and its low-energy eigenvalues and eigenvectors wer determined via direct diagonalization using the iterative Davidson algorithm~\cite{Crouzeix!994} across multiple classical computing nodes~\cite{robledomoreno2024chemistryexactsolutionsquantumcentric}.

To obtain the set of configurations, we sampled a suitable circuit ansatz on a quantum device. The Local Unitary Cluster Jastrow (LUCJ) \cite{Motta_2023} ansatz is a variational quantum circuit designed to generate approximate ground states of electronic systems while maintaining a moderate circuit depth. It is derived from unitary coupled-cluster (UCC) theory and incorporates local approximations to reduce the number of required two-qubit gates. The LUCJ wavefunction is constructed as
\begin{equation}
    |\Psi\rangle = \prod_{\mu=1}^{L} e^{\hat{K}_\mu} e^{i \hat{J}_\mu} e^{-\hat{K}_\mu} |x_{\text{\tiny RHF}}\rangle,
\end{equation}
where $\hat{K}_\mu = \sum_{pr,\sigma} K^\mu_{pr} \hat{a}^\dagger_{p\sigma} \hat{a}_{r\sigma}$ are one-body excitation operators, $\hat{J}_\mu = \sum_{pr,\sigma\tau} J^\mu_{p\sigma, r\tau} \hat{n}_{p\sigma} \hat{n}_{r\tau}$ are density-density interactions and $|x_{\text{\tiny RHF}}\rangle$ represents the reference Hartree-Fock (RHF) state encoded in the Jordan-Wigner mapping.

The circuit depth can be further reduced by employing a truncated version of the LUCJ ansatz. Specifically, we used the form
\begin{equation}
    |\Psi\rangle = e^{\hat{K}_2} e^{-\hat{K}_1} e^{i \hat{J}_1} e^{\hat{K}_1} |x_{\text{\tiny RHF}}\rangle.
\end{equation}
which represents to a two-layer LUCJ circuit, in which the final orbital rotation $e^{\hat{K}_2}$ and Jastrow interaction $e^{i \hat{J}_2}$ are omitted. This reduces the circuit depth to that of a single-layer LUCJ. \cite{robledomoreno2024chemistryexactsolutionsquantumcentric}

%The necessity of including \(e^{\hat{K}_2}\) in the truncated ansatz arises from the behavior of dynamically correlated systems. When setting ansatz parameters using the \(t_2\) amplitudes from a restricted closed-shell CCSD calculation, the \(J_\mu\) parameters can sometimes be small, causing \(e^{-\hat{K}_1}\) and \(e^{\hat{K}_1}\) to nearly cancel out. This results in a wavefunction that is overly concentrated around the HF configuration, limiting its ability to represent the true ground state. The presence of \(e^{\hat{K}_2}\) counteracts this effect by redistributing the wavefunction and improving the diversity of sampled configurations, leading to a more physically meaningful approximation of the correlated ground state. \cite{robledomoreno2024chemistryexactsolutionsquantumcentric}
 
We employed an extension of the SQD method, termed Extended SQD (Ext-SQD), to increase accuracy of molecular ground-state energies. Ext-SQD incorporates elements of Quantum Subspace Expansion while maintaining the computational efficiency of SQD. In contrast to standard SQD, which forms excited states by providing multiple eigenpairs of a projected Hamiltonian, Ext-SQD expands the accessible configuration space by explicitly applying excitation operators to the computational basis states in the original subspace.

% Ext-SQD also improves computational efficiency by selectively expanding the configuration space while discarding low-weight configurations based on a predefined amplitude threshold. Before applying excitation operators to extend the subspace, configurations with amplitudes below a certain cutoff are removed, reducing the total number of states without significantly affecting accuracy. The optimal threshold depends on the system being studied, ensuring a balance between computational cost and precision. Additionally, Ext-SQD allows control over which types of excitations—singles, doubles, triples, etc —are included, with accuracy and computational demands scaling accordingly. We can again systematically increase these until we are satisfied with the balance between cost and precision, which ensures that the expanded subspace remains both computationally feasible and sufficiently accurate for capturing essential correlation effects.

To efficiently sample from a variational ansatz, we adopted a classical simulation strategy based on truncated MPS, following reference [44]. Rather than using variational circuit compiling, we directly simulated structured quantum circuits, e.g., LUCJ, using the MPS backend of the Qiskit AerSimulator[63]. The output state was then approximated as an MPS with controlled entanglement, allowing us to avoid quantum hardware.

Unlike Ref.~\cite{Robertson_2025}, we did not employ a variational optimization routine or fidelity cost function to construct a new parameterized circuit. Instead, we relied on the intrinsic compressibility of the circuit output and the effectiveness of tensor network simulation for shallow-depth, structured circuits. The resulting MPS representations bypassed the need for quantum hardware execution and enabled scalable generation of bitstring samples from variational states for use in SQD post-processing.

In our PT2 calculations, we used a second-order perturbative correction of the Ext-SQD variational state to recover correlations from orbitals located outside the active space, following the Epstein–Nesbet form where external excitations provide the denominators and couplings to the active state provide the numerators. In standard multireference settings, this correction can be implemented with CASPT2 or NEVPT2, which construct denominators from a zeroth-order Hamiltonian and assemble numerators using one- and two-electron integrals jointly with reduced density matrices of the variational state \cite{Andersson1992,Angeli2001sc,Forsberg1997,Ghigo2004}. In our case, this procedure was straightforward as the Dice solver \cite{qiskit-addon-dice-solver} used in Ext-SQD supported NEVPT2 corrections.

\backmatter

% \bmhead{Supplementary information}
% If your article has accompanying supplementary file/s please state so here. 
% Please refer to Journal-level guidance for any specific requirements.

\bmhead{Acknowledgments}

We thank Ben Jaderberg, Edward Chen, Mario Motta, Mirko Armico, Gavin Jones,  Ieva Liepuoniute, Kevin Sung, Javier Robledo Moreno, Kunal Sharma and  Petar Jurcevic for discussions and technical assistance.

\section*{Declarations}

\bmhead{Code availability}
Open source-code repositories used in this work: PySCF \cite{pyscf}, \texttt{ffsim} \cite{ffsim},
Qiskit SDK \cite{qiskit}, \texttt{qiskit-addon-sqd} \cite{qiskit-addon-sqd}, \texttt{qiskit-addon-aqc-tensor} \cite{qiskit-addon-aqc-tensor}, mapomatic \cite{mapomatic} and Ray\cite{ray}. In addition, we used the Vienna Ab initio Simulation Package (VASP)\cite{Kresse1993,Kresse1996,Kresse1999}.

% Some journals require declarations to be submitted in a standardised format. Please check the Instructions for Authors of the journal to which you are submitting to see if you need to complete this section. If yes, your manuscript must contain the following sections under the heading `Declarations':

% \begin{itemize}
% \item Funding
% \item Conflict of interest/Competing interests (check journal-specific guidelines for which heading to use)
% \item Ethics approval and consent to participate
% \item Consent for publication
% \item Data availability 
% \item Materials availability
% \item Code availability 
% \item Author contribution
% \end{itemize}

% \noindent
% If any of the sections are not relevant to your manuscript, please include the heading and write `Not applicable' for that section. 

%%===================================================%%
%% For presentation purpose, we have included        %%
%% \bigskip command. Please ignore this.             %%
%%===================================================%%

\begin{appendices}
\section{SQD results}\label{sec:SQD_results}

\begin{figure}[!htb]
    \centering  % Aligns at top
    \includegraphics[width=\textwidth]{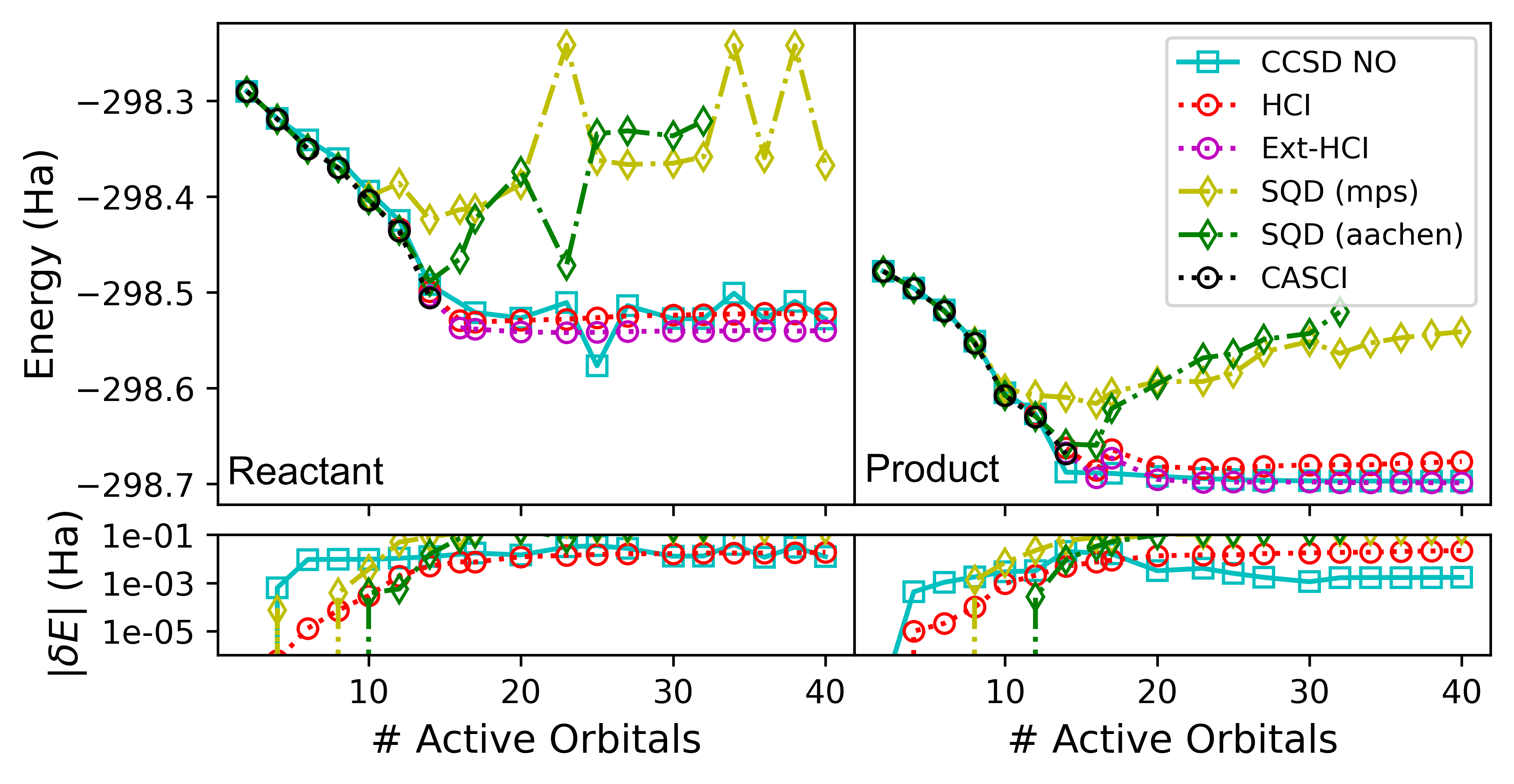}
    \caption{\textbf{Comparison of ground state and reaction energy results for SQD methods.} SQD calculations of the ground state energies as a function of active space size. The left curves are for the reactant and the right ones are for the product.}
    \label{fig:SQD_appendix}
\end{figure}

As Supplementary Information, we provide the SQD results for both reactant and product in \autoref{fig:SQD_appendix}.

% An appendix contains supplementary information that is not an essential part of the text itself but which may be helpful in providing a more comprehensive understanding of the research problem or it is information that is too cumbersome to be included in the body of the paper.

%%=============================================%%
%% For submissions to Nature Portfolio Journals %%
%% please use the heading ``Extended Data''.   %%
%%=============================================%%

%%=============================================================%%
%% Sample for another appendix section			       %%
%%=============================================================%%

%% \section{Example of another appendix section}\label{secA2}%
%% Appendices may be used for helpful, supporting or essential material that would otherwise 
%% clutter, break up or be distracting to the text. Appendices can consist of sections, figures, 
%% tables and equations etc.
\end{appendices}

%%===========================================================================================%%
%% If you are submitting to one of the Nature Portfolio journals, using the eJP submission   %%
%% system, please include the references within the manuscript file itself. You may do this  %%
%% by copying the reference list from your .bbl file, paste it into the main manuscript .tex %%
%% file, and delete the associated \verb+\bibliography+ commands.                            %%
%%===========================================================================================%%

\bibliography{sn-bibliography}% common bib file
%% if required, the content of .bbl file can be included here once bbl is generated
%%\input sn-article.bbl
\end{document}